\documentclass[aps,prl,10pt,preprint,groupedaddress]{revtex4-1}

\usepackage{graphics}
\usepackage{graphicx}
\usepackage{color}
\usepackage[ansinew]{inputenc}
\usepackage{amsmath}
\usepackage{amssymb}
\usepackage{color}
\usepackage{upgreek}
\usepackage{epsfig}

\begin{document}
\renewcommand{\tablename}{TAB.}

\title{Room-temperature condensation in whispering gallery microresonators assisted by longitudinal optical phonons}

\author{C.P. Dietrich$^{1,2}$}
\email{c.p.dietrich@tue.nl}
\author{R. Schmidt-Grund$^1$}
\email{schmidt-grund@physik.uni-leipzig.de}
\author{T. Michalsky$^1$}
\author{M. Lange$^1$}
\author{M. Grundmann$^1$}
\affiliation{$^1$Universit\"{a}t Leipzig, Institut f\"{u}r Experimentelle Physik II, Linn\'estr. 5, 04103 Leipzig, Germany}
\affiliation{$^2$SUPA, School of Physics and Astronomy, University of St Andrews, North Haugh, KY16 9SS St Andrews, United Kingdom}

\date{\today}

\begin{abstract}
We report condensation of hexagonal whispering gallery modes (WGM) at room temperature in ZnO microwires that embody nearly perfect polygonal whispering gallery microresonators. The condensate regime is achieved in the UV spectral range only at energies below the first longitudinal optical (LO) phonon replica of the free ZnO A-exciton transition and at non-zero wave vectors. We demonstrate that the multimodality of the WGM system and the high population of free excitons and phonons with various momenta strongly enhance the probability of an interaction of quasiparticles of the cavity exciton-photon system with LO phonons. We further examine the far-field mode pattern of lasing WGM and demonstrate their spatial coherence.
\end{abstract}

\maketitle

For more than 100 years whispering gallery mode (WGM) resonators\cite{Rayleigh1910} have attracted considerable interest in the scientific community. The striking supremacy of photonic WGM resonators among conventional Fabry-Pérot microcavities is manifested in their record mode quality factors ($Q$) and ultra-low lasing threshold\cite{Vahala2003} and is based on the nearly loss-free propagation of optical waves along the resonator circumference by total internal reflection. Besides that, the fabrication of WGM resonators is a lot less demanding compared to the fabrication of high-$Q$ Fabry-Pérot mode resonators which is always connnected with the need for highly reflecting mirrors that are obsolete for WGM resonators. The commonly used WGM resonator cross section is circular\cite{McCall1992}, but also polygonal structures entered the field in the last decade. Among them, hexagonal WGM as they naturally occur in wide bandgap semiconducting microwires (such as from ZnO\cite{Nobis2004} or GaN\cite{Coulon2012}) are remarkably unique since they combine high quality optical modes with the outstanding electronic properties of room-temperature stable excitons emitting in the UV spectral range. This also makes them promising candidates for the realization of polariton lasing in case of strong coupling between excitons and WGM photons\cite{Trichet2011}.

However, lasing of hexagonal WGM has elusively been demonstrated in ZnO structures\cite{Wang2006,Dai2009,Gargas2010,Chen2011,Dai2011,Dietrich2012}. The conspicuous similarity between all reported lasing spectra is their lasing wavelength that is always around 390\,nm (3.18\,eV), independent from the proposed gain mechanism of excitonic\cite{Versteegh2012} or polaritonic\cite{Xie2012} origin. This is surprising given the fact that the excitonic transitions in ZnO are a few tens of meV higher in energy and thus do not overlap with the observed spectral range for lasing.

In this letter, we report the investigation of lasing processes by stimulated mode gain in hexagonal ZnO microwires at room-temperature by means of reciprocal space Fourier spectroscopy. It is observed that the lasing only occurs at distinct, non-zero emission angles and at energies smaller than the transition of the first longitudinal optical (LO) phonon replica of the ZnO free A-excitons. We attribute this to strong interaction of WGM photons with LO phonons that ends up in excitonic states or lower photonic states under phonon energy and momentum transfer. This very efficient process seems to be supported by the high population of excitons at room temperature and the multimodality of the photonic WGM system which provides a wide possible momentum range for target states. In this context, we propose a condensation mechanism of either polaritonic or photonic origin to be responsible for the observed phenomena. Furthermore, we give new insight insight into the far-field behavior of phonon-assisted lasing processes and present double slit experiments proving the spatial coherence of the emission.

\begin{figure}
\includegraphics[width=0.66\textwidth]{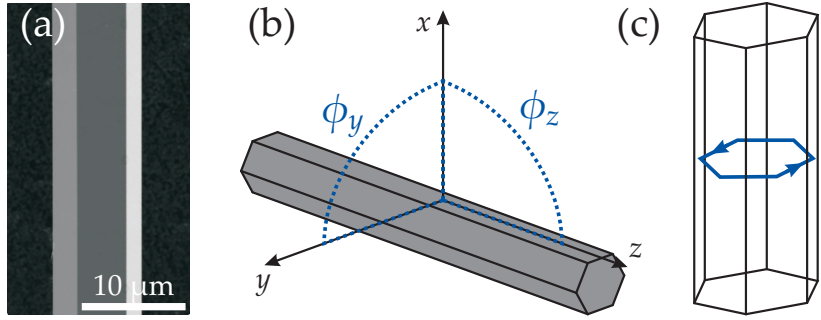}
\caption{\label{Fig1}(a) SEM picture of a hexagonal ZnO microwire with a diameter of around 9\,$\upmu$m. (b) Wire coordinate system: $z$-axis equals wire axis, $xy$-plane coincides with wire cross section. (c) Ray-optical pathway of WGM in a hexagonal resonator.}
\end{figure}

The investigated ZnO microwires were fabricated by a simple carbothermal vapor phase transport method for which powders of highly pure ZnO and carbon are pressed in a tablet shape and then put into a tube furnace. Optimal growth conditions are achieved at 1150$^\circ$C and ambient air conditions. The obtained ZnO microwires have nearly perfect hexagonal shape (see e.g. Fig.\ref{Fig1}a) and diameters (lenghts) in the $\upmu$m (mm) range\cite{Dietrich2012}. To simplify the coordination along the microwires, we define the wire axis as $z$-axis and the hexagonal wire cross section as $xy$-plane (illustrated in Fig.\ref{Fig1}b). According to this, $\phi_y$ ($\phi_z$) is the angle between the $x$- and $y$- ($z$-) direction. The wire luminescence was excited by a pulsed Thales DIVA II Nd:YAG UV laser with a wavelength of 266\,nm, pulse widths of 25\,ns and a repetition rate of 20\,Hz (excitation spot is elliptical with around 200\,$\upmu$m width). Light is collected in a micro-PL configuration by an objective (NA = 0.5, magnification $\Gamma$ = 100), spectrally dispersed by a grid spectrometer (320\,mm focal length, 2400 mm$^{-1}$ grating) and detected by a multi-channel back-illuminated CCD ($1024\times256$ pixels). For reciprocal (real) space Fourier imaging, we put two (three) additional lenses into the beam path at the positions of the Fourier planes (for details see \cite{Lai2008}).

\begin{figure}
\includegraphics[width=0.99\textwidth]{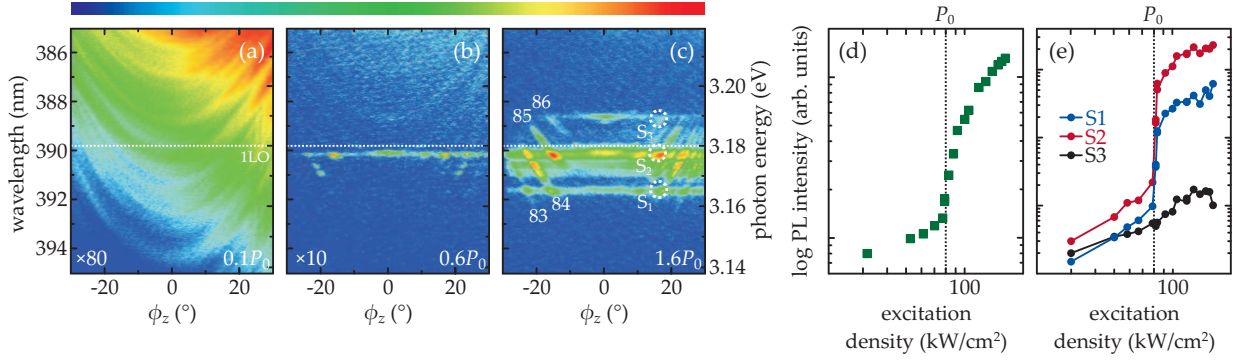}
\caption{\label{Fig2}(a-c) Excitation-dependent Fourier PL images of a ZnO microwire recorded at room temperature. The dashed white line marks the energy of the first LO-phonon replica of free ZnO A-excitons. The color-coded intensity scale is logarithmic, images are normalized to their intensity maximum. The intensity multiplication factors are given in lower left corners, the relative excitation densities in lower right corners. Numbers in (c) are mode numbers of lasing WGM. (d,e) Integrated PL intensity vs. excitation density for the total intensity (d) and partial intensity of spots S$_1$, S$_2$ and S$_3$ (e). The vertical dashed line highlights the lasing threshold at $P_0$ = 80\,kW/cm$^2$.}
\end{figure}

\begin{figure}
\includegraphics[width=0.75\textwidth]{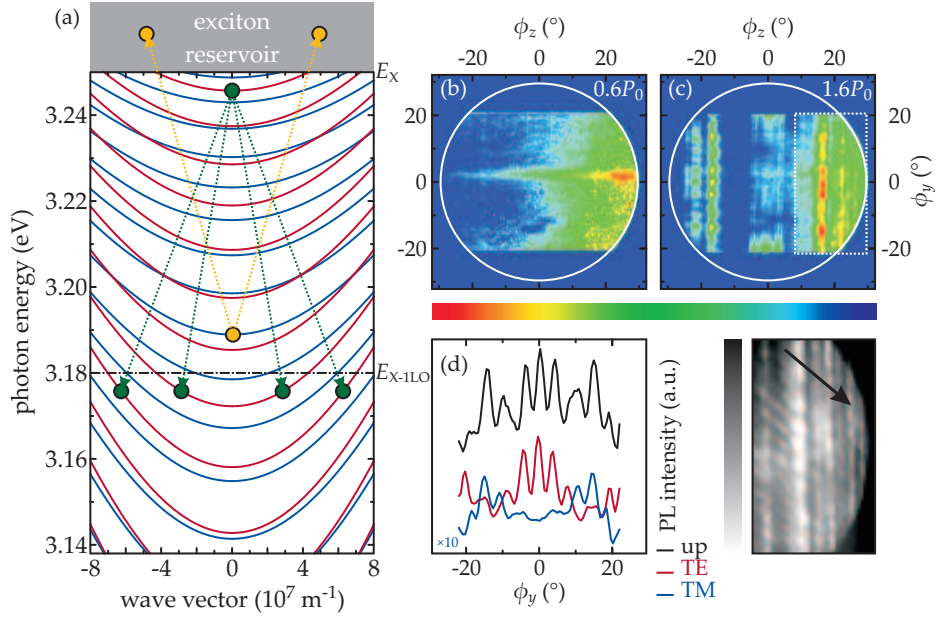}
\caption{\label{Fig3}(a) Phonon scattering scheme in a multimode whispering gallery mode system with exciton reservoir (grey area). Red and blue lines represent dispersion curves of TE- (red) and TM- (blue) polarized WGM. Yellow circles and dotted arrows highlight possible scattering paths of photons into excitonic states whereas green circles and dotted arrows show possible scattering into lower photonic states and the accompanied increae of the wave vector. (b,c) Fourier images of the complete reciprocal space for excitation below (b) and above (c) lasing threshold. (d) Intensity profile from (c) at $\phi_z = 18$º for unpolarized (up, black line), TE- (red) and TM- (blue) polarized PL.}
\end{figure}

Room-temperature Fourier photoluminescence (PL) images of a ZnO microwire recorded at different excitation densities are shown in Fig.\ref{Fig2}(a-c). The corresponding excitation-dependent, integrated PL intensity is depicted in Fig.\ref{Fig2}d. It cleary shows a threshold behavior at $P_0 = 80\,$kW/cm$^2$ and thus confirms the regime of laser-like emission for excitations above $P_0$. The respective relative excitation powers for Fig.\ref{Fig2}(a-c) are given in the lower right corners with respect to the lasing threshold. Far below threshold ($0.1P_0$), the Fourier image shows several parabola being distributed in energy and having a minimum at $\phi_z = 0$. These parabola are the dispersion curves of whispering gallery cavity modes present in the microwire. The dispersions can be approximated by $E(\phi) \propto \sin^2\phi$ and arise by the photonic confinement within the wire cross section\cite{Dietrich2011}. Note, that due to the optical anisotropy of ZnO both TE- and TM-polarized cavity modes are present in the spectra with slighty different dispersions (see also Fig.\ref{Fig3}a). Close to threshold ($0.6P_0$), these dispersion curves can still be observed but with much weaker intensity compared to individual spots appearing around 390\,nm. It is surprising that these spots do not appear at zero momentum as they would for regular photonic lasing, but have finite momentum and are symmetrically distributed around the dispersion minimum.

Another unexpected observation is a well-defined spectral range for the presence and absence of laser-like emission of the energetically distributed, individually emerging peaks. A clear spectral border between the lasing and linear regime is apparent for excitation power dependent integrated PL intensities of three different spots S$_1$, S$_2$ and S$_3$ marked by white, dashed circles in Fig.\ref{Fig2}c. These emission spots belong to WGM numbers 83, 84 and 85, respectively. Whereas peaks S$_1$ and S$_2$ with wavelengths above 390\,nm cleary cross the transition to the lasing regime, this is not the case for peak S$_3$ at a wavelength below 390\,nm although it is energetically closer to the exciton transitions. This "`border"' wavelength coincides with the energy of the first LO-phonon replica (LO phonon energy in ZnO is 72\,meV) of the free A-exciton transition in ZnO microwires around 3.25\,eV (the 1LO phonon transition is marked in Fig.\ref{Fig2}(a-c) by a horizontal white dashed line). Thus, only modes at energies (slightly) below the 1LO transition undergo the lasing transition. This indicates an effective interaction of WGM with phonons and may also explain the occuring lasing with non-zero momentum. An interaction of quasiparticles with phonons in ZnO was already reported for a planar microvcavity in the strong coupling regime where the polariton lasing threshold was lowest when the lower polariton branch spectrally overlapped with the 1LO phonon replica of the free excitons\cite{Orosz2012}. This phenomenon is similar to the strong phonon interaction in organic semiconductors\cite{Kena-Cohen2010}.

In general, the observed phonon processes can have different origin. In particular, they are most probably caused by either exciton-phonon (i), polariton-phonon (ii) or photon-phonon interactions (iii). In the following, we will discuss all three possible processes: (i) Exciton-phonon interaction (e.g. driven by an electron-hole plasma) has already extensively been studied in ZnO and is expected to occur more likely at the second phonon replica of the A-exciton transition\cite{Klingshirn1973}. Furthermore, an exciton-phonon process is not able to explain the discrete energetic distribution of lasing modes. For a broad excitonic band in interaction with phonons one would rather expect the lasing to occur along a major part of the dispersion instead of distinct, in energy separated lasing lines. Additionally, a red shift of the gain feeding the WGM lasing modes with increasing pumping power should be observable because it is provided by highly populated exciton states. Since we do not observe any indications for broad lasing bands or red shifting in our structures, we proceed on the assumption that this process is not responsible for the occuring lasing regime and neglect this in further discussions.

(ii) Besides lasing processes fed by exciton interaction, recent investigations on ZnO microwires also report the formation of exciton-polaritons that are created by the strong coupling between WGM photons and ZnO excitons (while WGM exciton polaritons are expected to have a high photonic component)\cite{Trichet2011,Xie2012}. The observed dispersion curves would then represent lower polariton branches and the observed transition to the lasing regime would coincide with the polariton condensation threshold. We emphasize that the observed lasing phenomena can indeed be explained by the polariton picture. In this context, a large number of initial polariton states distributed in energy and momentum would be available for the phonon scattering process into the target states with high momentum. Furthermore, it would also be reasonable for polariton lasing to occur below the 1LO transition because the bottlenecks of intital polaritons are also energetically located below $E_\text{X}$. This assumption is also supported by the one-dimensional character of the WGM polariton system that only allows propagation along the wire axis manifested in high WGM momenta and accompanied kinetic energies. Note that a small blueshift of lasing polaritons with increasing pump power would be expected since the highly populated polariton state should be subject to increased polariton-polariton scattering. Such a blueshift is absent in Fig.\ref{Fig2} but might also be smaller than the spectral resolution of our setup.

(iii) The scattering of photons with phonons (Raman-like process) is very unlikely and is only efficient in case cavity photons undergo stimulated scattering with phonons. This in turn is only possible if the scattering process is stimulated by a huge reservoir of photons, i.e. a photonic condensate. Such a condensate of photons could indeed be observed recently in dye-filled microcavities\cite{Klaers2010}. Even in common planar semiconductor Fabry-Pérot microcavities indications for photonic condensation were found by the built up of a spontaneous polarization in both the weak and strong coupling regime\cite{Ohadi2012,Kammann2012}. In the case of photonic condensation, the lasing states below the 1LO transition would then be fed by bosonic final state stimulation. That this assumption is reasonable can be supported by the absence of any blue- or redshift of the lasing peaks with increasing excitation density. One would expect that due to bosonic final state stimulation excitons are forced to decay in order to provide intial photons for the interaction with phonons. This process might cause a slight redshift of the excitonic state. However, since only the final state is fixed, phonons can randomly pick another state that fits both in energy and momentum. Therefore, it is unlikely to observe any spectral shifts for photonic condensation.

Based on the discussions above, we propose a possible phonon interaction scheme of WGM that is schematically shown in Fig.\ref{Fig3}a together with the calculated dispersions of TE- (red lines) and TM-polarized (blue lines) WGM in a hexagonal resonator with same diameter as the microwire from Fig.\ref{Fig2}. Here, $E_\text{X}$ denotes the energy of the A-excitonic states and $E_\text{X-1LO}$ is the energy of the phonon replica of the excitonic state. Assuming a photon emitted from the decay of the energetically broad exciton states in ZnO (at room temperature)\cite{Shan2005} or a respective polariton possesses an initial energy $E_\text{X} > E_\text{Ph,i} > E_\text{X-1LO}$, then it can undergo two possible scattering paths: (i) The photon/polariton can be scattered by a phonon into an excitonic state at higher energies by annihilation of the phonon (yellow dashed lines in Fig.\ref{Fig3}a). The photon/polariton then gains phonon energy and momentum to its final state $E_\text{Ph,f} > E_\text{X}$. This is possible since there is always a free exciton state with matching energy and momentum for the process. The wide range of possible excitonic momenta is thereby caused by the large exciton mass and its subsequent flat dispersion. The wide energetic range of available excitonic states (exciton band) is caused by the high density of states of A-, B- and C-excitons at room temperature and their broadened transitions. This process leads to the depletion of emission with photon/polariton energies above $E_\text{X-1LO}$. In this regard, it is also evident that there are still photonic/polaritonic modes observable close but above the 1LO phonon line (e.g. peak S3 in Fig.\ref{Fig2}c) since for these modes less excitonic target states are available at the low-energy edge of the excitonic band. (ii) The photon/polariton will be scattered into a photonic/polaritonic WGM state below the first LO phonon replica with $E_\text{Ph,f} < E_\text{X-1LO}$ by creation of a phonon (green dashed lines in Fig.\ref{Fig3}a). This relaxation process can only be vertical (without change in momentum) if the separation between initial and final state matches the LO phonon energy. Since this is only very rarely the case caused by the non-constant separation of WGM due to the strong refractive index dispersion in this spectral range\cite{Dietrich2011}, photons/polaritons will scatter into lower states by simultaneous gain of momentum as evidenced in Fig.\ref{Fig2}. Finally to complete the picture, it is necessary to mention that an initial state with $E_\text{Ph,i} < E_\text{X-1LO}$ will not have enough energy to be scattered by a phonon into an excitonic state and therefore will always end up in a photonic/polaritonic state.

As obvious from Fig.\ref{Fig2}, the phonon-assisted lasing emission with non-zero momentum strongly affects the directionality of the outcoupled light. As we already know from Fig.\ref{Fig2} that lasing only occurs at finite angles in the $xz$-plane, we recorded images of the full reciprocal space in order to also gain insight into the emission profile in the $xy$-plane. Respective images of the full reciprocal space including the $xz$-plane and the $xy$-plane are shown in Fig.\ref{Fig3}b,c for excitation below (b) and above (c) lasing threshold. Whereas there is hardly any pattern visible below threshold, clear discrete emission spots form for excitation above. Naturally, the lasing signatures along $\phi_z$ reflect the observations from Fig.\ref{Fig2}c - that is the absence of lasing peaks around zero momentum for all angles $\phi_y$ due to the involvement of phonons. Besides that, Fig.\ref{Fig3}c also shows a unique emission profile along $\phi_y$ which is shaped by several spots strung together for constant $\phi_z$. Since the reciprocal Fourier image mimics the far-field emission of the ZnO microwires, these spots can be interpretated as the far-field pattern of WGM involved in the particular stimulated emission process. The maxima then represent the nodes of the pattern that is different between WGM according to their mode number. Since spots at higher momenta arise from lasing along the dispersion curve of WGM with lower mode number, also the far-field pattern at higher momenta changes due to a decrease of lasing nodes (see magnified section of Fig.\ref{Fig3}c).

Note that the intensity of the lasing peaks is quite uniform along the microwire circumference which is caused by the superposition of TE- and TM-polarized modes, see Fig.\ref{Fig3}d. Here, TE-polarized WGM preferentially couple out along the hexagon corners and facets ($\pi$/6 period) whereas most promiment TM-polarized modes are found rotated by $\pi$/12. In this regard, the observed far-field patterns agree well with the photonic condensation and polariton interpretation since WGM exciton-polaritons are also expected to be of very high photonic character.

\begin{figure}
\includegraphics[width=0.75\textwidth]{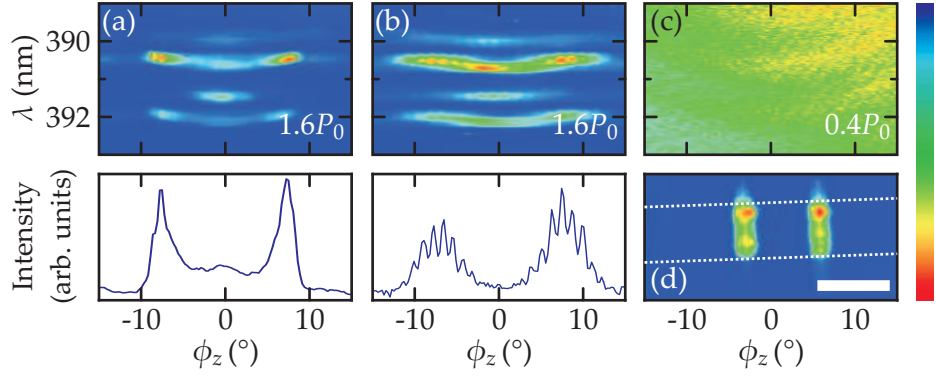}
\caption{\label{Fig4}(a-c) Fourier PL images of a ZnO microwire recorded at room temperature after passing a single (a) or double slit (b,c) above (a,b) or below (c) lasing threshold. Below (a,b): Intensity profiles from images above for wavelength $\lambda = 390.5$\,nm. (d) Real space PL image of a ZnO microwire when passing the double slit. The length of the scale bar represents 10\,$\upmu$m. The white dashed line marks the wire edges.}
\end{figure}

In order to examine the spatial coherence of the stimulated emission, we performed experiments with both single and double slits\cite{Deng2007}. More precisely, we put slits at the position of the real space Fourier plane (see real space image in Fig.\ref{Fig4}d for the double slit) into the detection beam path before the spectrometer and record the reciprocal space Fourier images as shown in Fig.\ref{Fig4}. The width of the slits is 2.5\,mm and the distance of double slits is 10\,mm. For a fully coherent light source, we would expect to observe interference fringes after passing a double slit due to a fixed phase relationship between photon wave packets. In Fig.\ref{Fig4}a, after passing a single slit, the microwire luminescence is similar to what was observed in Fig.\ref{Fig2}c. However, the situation changes for the double slit. Below threshold (Fig.\ref{Fig4}c), the wire luminescence does not show any significant intensity modulation except that due to photonic dispersions. However, above lasing threshold clear fringes can be observed. See also Fig.\ref{Fig4}b above for comparison showing the intensity at 390\,nm vs. emission angle. Herewith, we clearly demonstrate the built-up of off-diagonal long-range spatial coherence in the WGM system when reaching the condensation regime.

In conclusion, we have investigated the room-temperature condensation of hexagonal WGM in ZnO microwires and observed a strong interaction with phonons that is evidenced by a pronounced momentum gain of condensed WGM and an obvious spectral limit for the condensation regime that matches the first phonon replica of the ZnO free A-exciton transition. A respective phonon scattering scheme of photons into either energetically higher excitonic or lower photonic states could be proposed. We further discussed the origin of the phonon-assisted lasing and conclude that electron-phonon interaction can be excluded here whereas bosonic condensation is responsible being either of photonic or polaritonic origin. Besides, a full record of the reciprocal space gave insight into the unique emission properties of WGM in phonon-assisted condensation processes and demonstrated the strong polarization-dependent directionality of WGM along the wire circumference. Finally, spatial coherence of condensed WGM in ZnO was proven by performing Young's double slit experiment.

This work has been supported by Leipzig School of Natural Sciences BuildMoNa (GS 185/1), the European Social Fund (ESF) within Nachwuchsforschergruppe 'Multiscale functional structures' and DFG-FOR1616 (SCHM2710/2-1,P1) "Dynamics and Interactions of Semiconductor Nanowires for Optoelectronics". We gratefully acknowledge target preparation by G. Ramm.

\end{document}